\documentclass[12pt]{article}
\usepackage{epsfig}
\def\be{\begin{equation}}
\def\ee{\end{equation}}
\def\bea{\begin{eqnarray}}
\def\eea{\end{eqnarray}}
\usepackage{graphicx}% Include figure files

\catcode`\@=11
\def\lsim{\mathrel{\mathpalette\@versim<}}
\def\gsim{\mathrel{\mathpalette\@versim>}}
\def\@versim#1#2{\vcenter{\offinterlineskip
\ialign{$\m@th#1\hfil##\hfil$\crcr#2\crcr\sim\crcr } }}
\catcode`\@=12
\usepackage{axodraw}

\catcode`@=12
\begin{document}
\thispagestyle{empty}
\begin{flushright}
UCRHEP-T540\\
February 2014\
\end{flushright}
\vspace{0.6in}
\begin{center}
{\LARGE \bf The Higgs Connection --\\ 
Flavor and Dark Matter\\}
\vspace{1.2in}
{\bf Ernest Ma\\}
\vspace{0.2in}
{\sl Department of Physics and Astronomy, University of California,\\
Riverside, California 92521, USA\\}
\end{center}
\vspace{0.8in}

\centerline{\bf \underline{Three families of quarks and leptons,}}
\vspace{0.2in}
\centerline{\bf \underline{one Higgs to rule them all,}}
\vspace{0.2in}
\centerline{\bf \underline{and in the darkness bind them.}}

\vspace{1.5in}
\noindent Talk at the ``International Conference on Flavor Physics and Mass 
Generation,'' Institute of Advanced Studies, Nanyang Technological 
University, Singapore (February 2014).

\newpage
\baselineskip 24pt

With the recent discovery~\cite{atlas12,cms12} of the 126 GeV particle at 
the Large Hadron Collider (LHC), and the likelihood of it being the one 
physical neutral Higgs boson $h$ of the standard model (SM) of quarks and 
leptons, it may be a good time to reflect on what it means as to some 
of the other outstanding problems in particle physics.  In this talk, I 
focus on flavor and dark matter (DM).  I will show in eight easy steps how 
the one Higgs boson may be the key to understanding flavor through its 
interaction with dark matter~\cite{m13}.

\noindent {\bf Step 1}: In addition to the observable sector of SM particles, 
let there be a dark sector, odd under $Z_2$ which may be derived from an 
$U(1)_D$ gauge symmetry.  The particles of this dark sector consist of 
three neutral singlet Dirac fermions $N_{1,2,3}$ and the scalars 
\begin{equation}
(\eta^+,\eta^0), \zeta^{-1/3} \sim \underline{5}, ~~~~~ 
(\xi^{2/3},\xi^{-1/3}), (\zeta^{2/3})^*, \chi^+ \sim \underline{10},
\end{equation}
which are complete $SU(5)$ multiplets.  The lightest $N$ is stable and 
a possible candidate for the observed dark matter of the Universe.

\noindent {\bf Step 2}:  A non-Abelian discrete symmetry is imposed, under 
which $N_{1,2,3}$ as well as the three families of quarks and leptons of 
the SM transform nontrivially.  As a 
concrete example, consider the non-Abelian discrete symmetry 
$A_4$~\cite{mr01,m02,bmv03,m04}, which is also the symmetry group of 
the tetrahedron.  It has four irreducible representations 
$\underline{1}, \underline{1}', \underline{1}'', \underline{3}$, 
with the multiplication rule
\begin{equation}
\underline{3} \times \underline{3} = \underline{1} + \underline{1}' + 
\underline{1}'' +  \underline{3} + \underline{3}.
\end{equation}

\noindent {\bf Step 3}: Let the leptons $L_{iL} = (\nu, l)_{iL} \sim 
\underline{1}, \underline{1}', \underline{1}''$, $l_{iR} \sim \underline{3}$, 
with $\Phi = (\phi^+,\phi^0) \sim \underline{1}$, 
then the usual SM Yukawa couplings $\bar{L}_{iL} l_{jR} \Phi$ are forbidden.  
Similarly, the quarks $Q_{iL} = (u,d)_{iL} \sim \underline{1}, \underline{1}', 
\underline{1}''$, $u_{iR} \sim \underline{3}$, $d_{iR} \sim \underline{3}$, 
thus also forbidding $\bar{Q}_{iL} d_{jR} \Phi$ and $\bar{Q}_{iL} u_{jR} 
\tilde{\Phi}$, where $\tilde{\Phi} = (\bar{\phi}^0,-\phi^-)$.
The nonzero vacuum expectation value of $\phi^0$ generates $W$ and $Z$ 
masses but not fermion masses.

\noindent {\bf Step 4}: Consider first the charged leptons with 
\begin{equation}
(\eta^+,\eta^0), ~\chi^+ \sim \underline{1}, ~~~ N_{iL} \sim \underline{3}, 
~~~ N_{iR} \sim \underline{1}, \underline{1}', \underline{1}''.
\end{equation}
Hence the Yukawa couplings $\bar{N}_{R} l_L \eta^+$ and $\bar{l}_R N_L \chi^-$ 
are allowed.  The soft breaking of $A_4$ to $Z_3$~\cite{m10} in the 
$3 \times 3$ Dirac mass matrix of $N_{1,2,3}$, i.e.
\begin{equation}
{\cal M}_N = 
{1 \over \sqrt{3}} \pmatrix{1 & 1 & 1 \cr 1 & \omega & \omega^2 \cr 
1 & \omega^2 & \omega} \pmatrix{M_1 & 0 & 0 \cr 0 & M_2 & 0 \cr 0 & 0 & M_3} ,
\end{equation}
where $\omega = \exp(2 \pi i/3)$, then allows $\Phi$ 
to couple to $\bar{l}_L l_R$ in one loop as shown in Fig.~1.  
\begin{figure}[htb]
\vspace*{-3cm}
\hspace*{-3cm}
\includegraphics[scale=1.0]{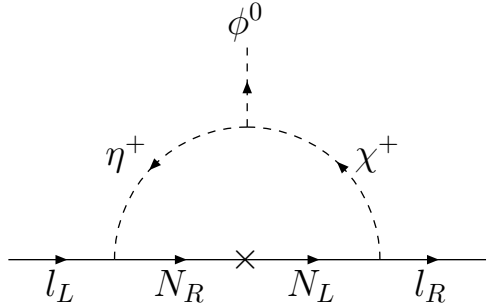}
\vspace*{-21.5cm}
\caption{One-loop generation of charged-lepton mass.}
\end{figure}

Similar diagrams exist for the quarks, using the other scalars of Eq.~(1).
\begin{figure}[htb]
\vspace*{-3cm}
\hspace*{-3cm}
\includegraphics[scale=1.0]{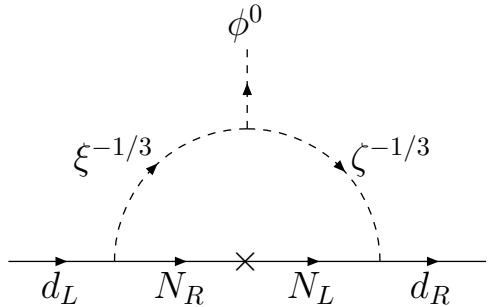}
\vspace*{-21.5cm}
\caption{One-loop generation of $d$ quark mass.}
\end{figure}
\begin{figure}[htb]
\vspace*{-3cm}
\hspace*{-3cm}
\includegraphics[scale=1.0]{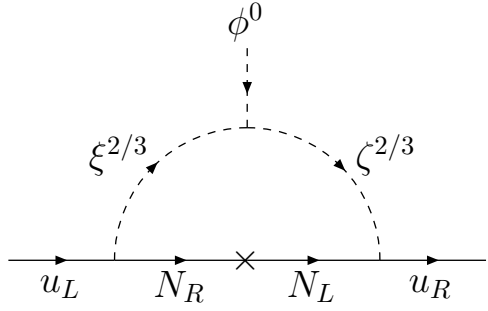}
\vspace*{-21.5cm}
\caption{One-loop generation of $u$ quark mass.}
\end{figure}

Thus all quarks and leptons owe their masses to dark matter in conjunction 
with $\Phi$.  Note that the DM particles are distinguished from the SM 
particles by $Z_2$ which may itself be a remnant of an $U(1)_D$ gauge 
symmetry.

\noindent {\bf Step 5}: The residual $Z_3$ symmetry maintains diagonal mass 
matrices for $u$ and $d$ quarks as well as charged leptons. This is an 
explanation of why the quark mixing matrix $V_{CKM}$ is 
nearly diagonal.

\noindent {\bf Step 6}:  Further soft breaking of $Z_3$ allows a realistic 
quark mixing matrix ($V_{CKM}$). The $\bar{q}_{iL} q_{jR}$ mass matrix is of 
the form
\begin{equation}
{\cal M}_q = \pmatrix{f_1 & 0 & 0 \cr 0 & f_2 & 0 \cr 0 & 0 & f_3} 
U_M^L \pmatrix{M_1 & 0 & 0 \cr 0 & M_2 & 0 \cr 0 & 0 & M_3} (U_M^R)^\dagger.
\end{equation}
If the unitary matrices $U_M^{L,R}$ are the identity, then $Z_3$ is not broken 
and $V_{CKM}$ is also the identity.  Let $U_M^L$ be 
approximately given by
\begin{equation}
U_M^L \simeq \pmatrix{1 & -\epsilon_{12} & -\epsilon_{13} \cr \epsilon_{12}^* 
& 1 & -\epsilon_{23} \cr \epsilon_{13}^* & \epsilon_{23}^* & 1}
\end{equation}
then 
\begin{equation}
{\cal M}_q {\cal M}_q^\dagger \simeq \pmatrix{f_1^2 M_1^2 & f_1 f_2 \epsilon_{12} 
(M_1^2 - M_2^2) & f_1 f_3 \epsilon_{13} (M_1^2 - M_3^2) \cr f_1 f_2  \epsilon_{12}^* 
(M_1^2 - M_2^2) & f_2^2 M_2^2 & f_2 f_3 \epsilon_{23} (M_2^2 - M_3^2) \cr 
f_1 f_3 \epsilon_{13}^* (M_1^2 - M_3^2) & f_2 f_3 \epsilon_{23}^* (M_2^2 - M_3^2) 
& f_3^2 M_3^2}.
\end{equation}
Let $m_d \simeq f_1^d M_1$, $m_s \simeq f_2^d M_2$, $m_b \simeq f_3^d M_3$, 
$m_u \simeq f_1^u M_1$, $m_c \simeq f_2^u M_2$, $m_t \simeq f_3^u M_3$, 
then $V_{CKM}$ is approximately 
given by
\begin{eqnarray}
&& V_{ud} \simeq V_{cs} \simeq V_{tb} \simeq 1, ~~~ 
V_{us} \simeq \left( {m_d \over m_s} - {m_u \over m_c} \right) \epsilon_{12} 
\left( {M_2^2 - M_1^2 \over M_2 M_1} \right), \\
&& V_{ub} \simeq \left( {m_d \over m_b} - {m_u \over m_t} \right) \epsilon_{13} 
\left( {M_3^2 - M_1^2 \over M_3 M_1} \right), ~~~ 
V_{cb} \simeq \left( {m_s \over m_b} - {m_c \over m_t} \right) \epsilon_{23} 
\left( {M_3^2 - M_2^2 \over M_3 M_2} \right).
\end{eqnarray}
There are many realistic solutions of the above.  The simplest is to 
set $f_1^d = f_2^d = f_3^d$, in which case $V_{CKM} \simeq (U_M^L)^\dagger$.  
In other words, the soft breaking of $Z_3$ which generates $U_M^L$ is directly 
linked to the observed $V_{CKM}$ .

\noindent {\bf Step 7}:  As for the radiative generation of Majorana 
neutrino mass, the well-studied one-loop scotogenic model~\cite{m06} 
(with $Z_2$) may be used.  If $U(1)_D$ is desired, then the recent 
one-loop proposal~\cite{mpr13} with two scalar doublets 
$(\eta_{1,2}^+, \eta_{1,2}^0)$ transforming oppositely under $U(1)_D$ is a 
good simple choice.  However, a two-loop realization may also be adopted, 
as shown in Fig.~4, which may preserve $U(1)_D$ as well.  Under $Z_3$, 
$\nu_{e,\mu,\tau}, N_{e,\mu,\tau}, \rho_{1,2,3} \sim 1,\omega,\omega^2$, 
$(\phi^+,\phi^0), (\eta^+,\eta^0), \chi^0 \sim 1$.  Under $U(1)_D$, $N, 
(\eta^+,\eta^0), \chi^0 \sim 1$, $\rho \sim 2$.
\begin{figure}[htb]
\vspace*{-3cm}
\hspace*{-3cm}
\includegraphics[scale=1.0]{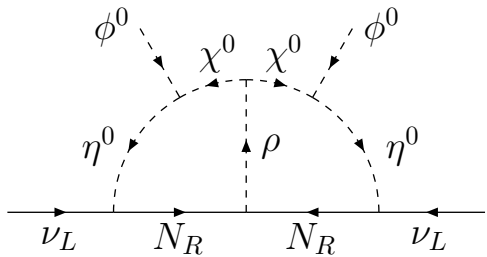}
\vspace*{-21.5cm}
\caption{Two-loop generation of Majorana neutrino mass with $U(1)_D$ symmetry.}
\end{figure}

The addition of $\chi^0$ and $\rho_1$ completes the two loops without 
breaking $U(1)_D$ or $Z_3$.  This would result in a Majorana neutrino 
mass matrix in the basis $(\nu_e,\nu_\mu,\nu_\tau)$ of the form
\begin{equation}
{\cal M}_\nu = \pmatrix{A & 0 & 0 \cr 0 & 0 & B \cr 0 & B & 0}.
\end{equation}
The further addition of $\rho_{2,3}$ together with the soft breaking of $Z_3$ 
using the trilinear $\chi^0 \chi^0 \rho_{2,3}^\dagger$ couplings allows 
${\cal M}_\nu$ to become
\begin{equation}
{\cal M}_\nu = \pmatrix{A & C  & C^* \cr C & D^* & B \cr C^* & B & D},
\end{equation}
where $A$ and $B$ are real.  Note that this pattern is protected by a 
symmetry first pointed out in Ref.~\cite{gl04}, i.e. $e \to e$ and 
$\mu - \tau$ interchange with $CP$ conjugation, 
and obtained previously in Ref.~\cite{bmv03}.  As such, it is also 
guaranteed to yield maximal $\nu_\mu - \nu_\tau$ mixing ($\theta_{23} = \pi/4$) 
and maximal $CP$ violation, i.e. $\exp(-{i \delta}) = \pm i$, whereas 
$\theta_{13}$ may be nonzero and arbitrary.

The mass matrix of Eq.~(11) has six parameters: $A, B, C_R, C_I, D_R, D_I$, but 
only five are independent because the relative phase of $C$ and $D$ is 
unobservable.  Using the conventional parametrization of the neutrino 
mixing matrix, the angle $\theta_{13}$ is given by
\begin{equation}
{s_{13} \over c_{13}} = {-D_I \over \sqrt{2} C_R}, ~~~~~~ 
{s_{13} c_{13} \over c_{13}^2 - s_{13}^2} = {\sqrt{2} C_I \over A - B + D_R}.
\end{equation}
The adjustable relative phase of $C$ and $D$ is used to allow the above 
two equations to be satisfied with a single $\theta_{13}$.  The angle 
$\theta_{12}$ is then 
given by
\begin{equation}
{s_{12} c_{12} \over c_{12}^2 - s_{12}^2} = {-\sqrt{2} (c_{13}^2 - s_{13}^2) C_R 
\over c_{13} [ c_{13}^2 (A - B - D_R) + 2 s_{13}^2 D_R]}.
\end{equation}
As a result, the three mass eigenvalues are 
\begin{eqnarray}
m_1 &=& {c_{13}^2 [c_{12}^2 A - s_{12}^2 B - s_{12}^2 D_R] - s_{13}^2 
[(c_{12}^2 - s_{12}^2) B - D_R] \over (c_{13}^2 - s_{13}^2)
(c_{12}^2 - s_{12}^2)}, \\ 
m_2 &=& {c_{13}^2 [-s_{12}^2 A + c_{12}^2 B + c_{12}^2 D_R] - s_{13}^2 
[(c_{12}^2 - s_{12}^2) B + D_R] \over (c_{13}^2 - s_{13}^2)
(c_{12}^2 - s_{12}^2)}, \\ 
m_3 &=& {s_{13}^2 A - c_{13}^2 B + c_{13}^2 D_R \over c_{13}^2 - s_{13}^2}.
\end{eqnarray}
Since $s_{13}^2 \simeq 0.025$ is small, these expressions become
\begin{eqnarray}
m_2 + m_1 &\simeq& A + B + D_R + s_{13}^2 (A - B + D_R), \\ 
(c_{12}^2 - s_{12}^2) (m_2 - m_1) &\simeq& -A + B + D_R - s_{13}^2 (A - B + D_R), \\ 
m_3 &\simeq& -B + D_R + s_{13}^2 (A - B + D_R).
\end{eqnarray}
It is clear that a realistic neutrino mass spectrum with $m_2^2 - m_1^2 << 
|m_3^2 - (m_2^2 + m_1^2)/2|$ may be obtained with either 
$|m_1| < |m_2| < |m_3|$ (normal ordering) or $|m_3| < |m_1| < |m_2|$ 
(inverted ordering).

\noindent {\bf Step 8}: The predicted scalars of Eq.~(1) which connect 
the quarks and leptons to their common dark-matter antecedents, i.e. 
$N_{1,2,3}$, are possibly observable at the LHC.  They may also change 
significantly the SM couplings of $\Phi$~\cite{fm13}.
In Fig.~1, $\eta^+$ is part of an electroweak doublet $(\eta^+,\eta^0)$ and 
$\chi^+$ is a singlet.  They mix because of the $\mu (\eta^+ \phi^0 - \eta^0 
\phi^+) \chi^-$ trilinear interaction.  The $2 \times 2$ mass-squared 
matrix spanning $(\eta^\pm,\chi^\pm)$ is given by
\begin{eqnarray}
{\cal M}^2_{\eta \chi} = \pmatrix{m^2_\eta & \mu v/\sqrt{2} \cr 
\mu v/\sqrt{2} & m^2_\chi} = \pmatrix{\cos \theta & -\sin \theta 
\cr \sin \theta & \cos \theta} \pmatrix{m_1^2 & 0 \cr 0 & m_2^2} 
\pmatrix{\cos \theta & \sin \theta \cr -\sin \theta & \cos \theta},
\end{eqnarray}
where $\langle \phi^0 \rangle = v/\sqrt{2}$.  Now  
$\zeta_1 = \eta \cos \theta + \chi \sin \theta$,  
$\zeta_2 = \chi \cos \theta - \eta \sin \theta$ 
are the mass eigenstates, and the mixing angle $\theta$ is given by
\begin{equation}
{\mu v \over \sqrt{2}} = \sin \theta \cos \theta (m_1^2 - m_2^2).
\end{equation}
The exact calculation of $m_l$ in terms of the exchange of $\zeta_{1,2}$ 
results in~\cite{m06,mpr13}
\begin{eqnarray}
m_l &=& {f_\eta f_\chi \sin \theta \cos \theta ~m_N \over 16 \pi^2} 
\left[ {m_1^2 \over m_1^2 - m_N^2} \ln {m_1^2 \over m_N^2} - 
{m_2^2 \over m_2^2 - m_N^2} \ln {m_2^2 \over m_N^2} \right] \nonumber \\ 
&=& {f_\eta f_\chi \mu  v ~m_N \over 16 \sqrt{2} \pi^2 (m_1^2 - m_2^2)} 
\left[ {m_1^2 \over m_1^2 - m_N^2} \ln {m_1^2 \over m_N^2} - 
{m_2^2 \over m_2^2 - m_N^2} \ln {m_2^2 \over m_N^2} \right].
\end{eqnarray}

Let $\phi^0 = (v + h)/\sqrt{2}$ and consider the effective Yukawa 
coupling $h \bar{l} l$.  In the SM, it is of course equal to $m_l/v$, 
but here it has three contributions.  Assuming that $m_h^2$ is small 
compared to $m_{1,2}^2$ and $m_N^2$, and defining $x_{1,2} = m_{1,2}^2/m_N^2$, 
Fig.~1 yields
\begin{eqnarray}
f_l^{(3)} = {f_\eta f_\chi \mu \over 16 \sqrt{2} \pi^2 m_N} 
[ (\cos^4 \theta + \sin^4 \theta) F(x_1,x_2) + \sin^2 \theta \cos^2 \theta 
(F(x_1,x_1) + F(x_2,x_2))], 
\end{eqnarray}
\begin{eqnarray}
F(x_1,x_2) = {1 \over x_1 - x_2} \left[ {x_1 \over x_1 -1} \ln x_1 - 
{x_2 \over x_2 -1} \ln x_2 \right],~~~F(x,x) = {1 \over x-1} - 
{\ln x \over (x-1)^2}.
\end{eqnarray}
Comparing Eq.~(23) with Eq.~(22), we see that $f_l^{(3)} = m_l/v$ only 
in the limit $\theta \to 0$.  We see also that $F(x_1,x_1)+F(x_2,x_2)$ is 
always greater than $2F(x_1,x_2)$, so that $f_l^{(3)}$ is always greater 
than $m_l/v$. The correction due to nonzero $m_h$ is easily computed 
in the limit $m_1=m_2=m_N$, in which case it is $m_h^2/12m_N^2$.  This 
shows that it should be generally negligible. Let
\begin{equation}
F_+(x_1,x_2) = {F(x_1,x_1) + F(x_2,x_2) \over 2 F(x_1,x_2)} - 1, ~~~ 
F_-(x_1,x_2) = {F(x_1,x_1) - F(x_2,x_2) \over 2 F(x_1,x_2)},
\end{equation}
then $F_+ \geq 0$ and if $x_1=x_2$, $F_+ = F_- = 0$.  
The other two contributions to $h \bar{\tau} \tau$ come from 
$\lambda_\eta v \eta^+ \eta^-$ and $\lambda_\chi v \chi^+ \chi^-$, i.e.
\begin{eqnarray} 
f^{(1)}_\tau &=& {\lambda_\eta v f_\eta f_\chi \over 
16 \pi^2 m_N} \sin \theta \cos \theta [\cos^2 \theta F(x_1,x_1)  - 
\sin^2 \theta F(x_2,x_2) - \cos 2 \theta F(x_1,x_2)], \\ 
f^{(2)}_\tau &=& {\lambda_\chi v f_\eta f_\chi \over 
16 \pi^2 m_N} \sin \theta \cos \theta [\sin^2 \theta F(x_1,x_1) 
- \cos^2 \theta F(x_2,x_2) + \cos 2 \theta  F(x_1,x_2)]. 
\end{eqnarray}
\begin{figure}[htb]
%\vspace*{-0.5cm}
\hspace*{2cm}
\includegraphics[scale=1.0]{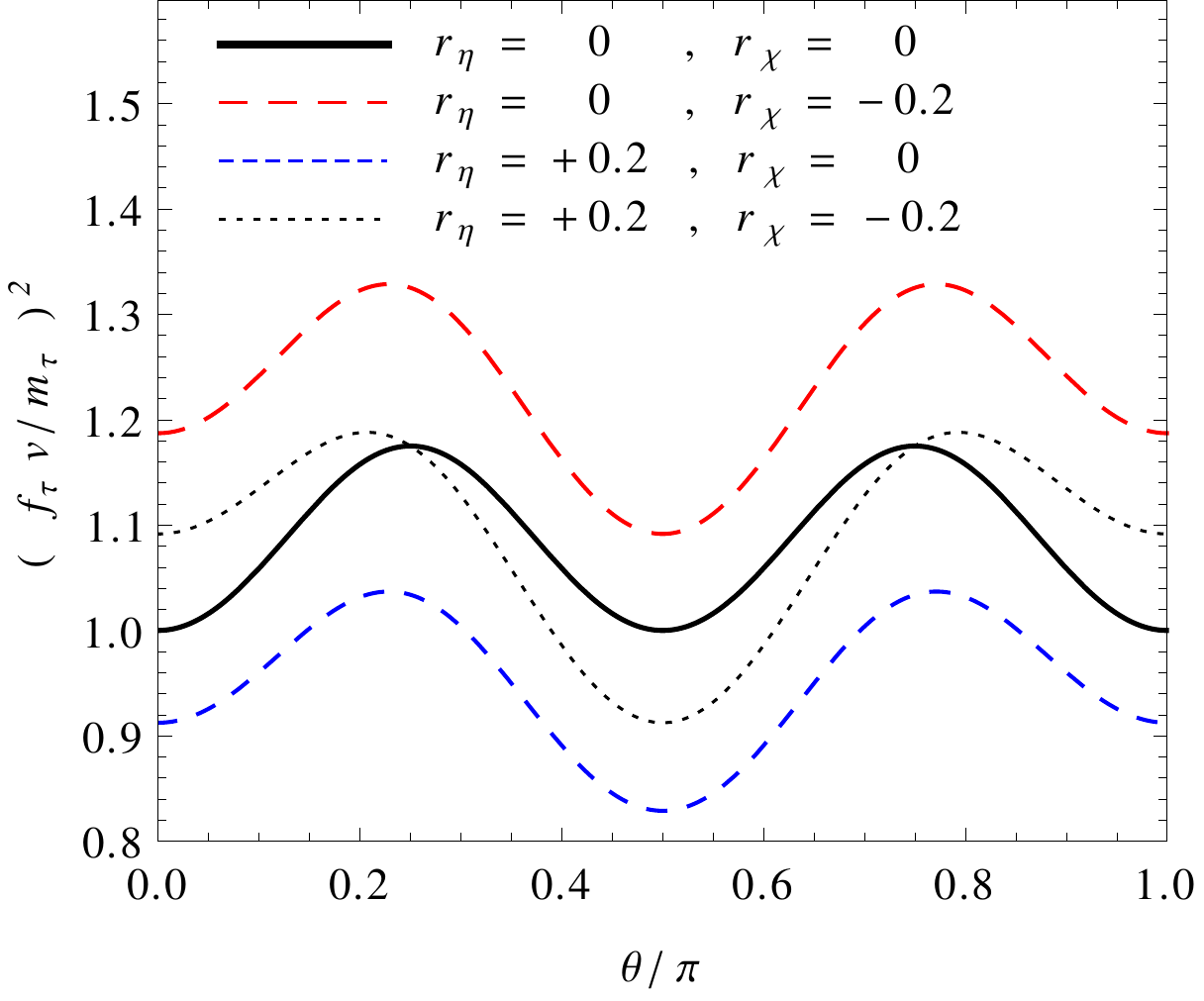}
\vspace*{-0.5cm}
\caption{The ratio $(f_\tau v/m_\tau)^2$ plotted against $\theta$ for $x=2$ 
and various $r_\eta$ and $r_\chi$.}
\end{figure}
Let $r_{\eta,\chi} = \lambda_{\eta,\chi} v^2 / m_N^2$, then the total 
contribution to the Higgs Yukawa coupling is given by
\begin{eqnarray}
{f_\tau v \over m_\tau} = 1 + \left[ {1 \over 2} (\sin 2 \theta)^2 + 
{1 \over 4} \sin 4 \theta (r_\eta - r_\chi) \right] F_+ + {1 \over 2} 
\sin 2 \theta (r_\eta + r_\chi) F_-.
\end{eqnarray}
The LHC measurements of $h \to \tau^+ \tau^-$ and 
$h \to \bar{b} b$ provide the bounds 
\begin{eqnarray}
&& \left( {f_\tau v \over m_\tau} \right)^2 = 1.4 \pmatrix{+0.5 \cr -0.4}, ~~~ 
\left( {f_b v \over m_b} \right)^2 = 0.2 \pmatrix{+0.7 \cr -0.6} 
~{\rm (ATLAS)}~\cite{atlas}, \\ 
&& \left( {f_\tau v \over m_\tau} \right)^2 = 1.10 \pm 0.41, ~~~
\left( {f_b v \over m_b} \right)^2 = 1.15 \pm 0.62~~
{\rm (CMS)}~\cite{cms}.
\end{eqnarray}
The ratio $(f_\tau v/m_\tau)^2$ is plotted against $\theta$ for $x=3$ 
and various $r_\eta$ and $r_\chi$ in Fig.~5.

%\newpage
In conclusion, the 126 GeV particle may hold secrets of physics beyond the 
SM.  It could indeed be the \underline{one} Higgs, but not exactly that 
of the SM.  Its couplings to fermions may hold the key to understanding 
flavor and dark matter.  Particles which look like scalar quarks and 
leptons are also predicted at the LHC, but with properties not exactly 
like those required by supersymmetry.  The Higgs Yukawa couplings to 
fermions may also differ significantly from the SM predictions.

\bigskip

I thank K. K. Phua and everyone at IAS/NTU for their great hospitality, and 
H. Fritzsch for organizing this conference.  My work is supported in part 
by the U.~S.~Department of Energy under Grant No.~DE-SC0008541.

\bibliographystyle{unsrt}

\end{document}